# Physical Interpretation of the 26 dimensions of Bosonic String Theory

February-October 2001; July 2002
Frank D. (Tony) Smith, Jr.
tsmith@innerx.net
http://www.innerx.net/personal/tsmith/Rzeta.html#bosonstring

Abstract:

The 26 dimensions of Closed Unoriented Bosonic String Theory
are interpreted as the 26 dimensions of
the traceless Jordan algebra J3(O)o of 3x3 Octonionic matrices,
with each of the 3 Octonionic dimenisons of J3(O)o
having the following physical interpretation:

4-dimensional physical spacetime plus 4-dimensional internal symmetry space;
8 first-generation fermion particles;
8 first-generation fermion anti-particles.

This interpretation is consistent with interpreting the strings
as World Lines of the Worlds of Many-Worlds Quantum Theory
and the 26 dimensions as the degrees of freedom of
the Worlds of the Many-Worlds.

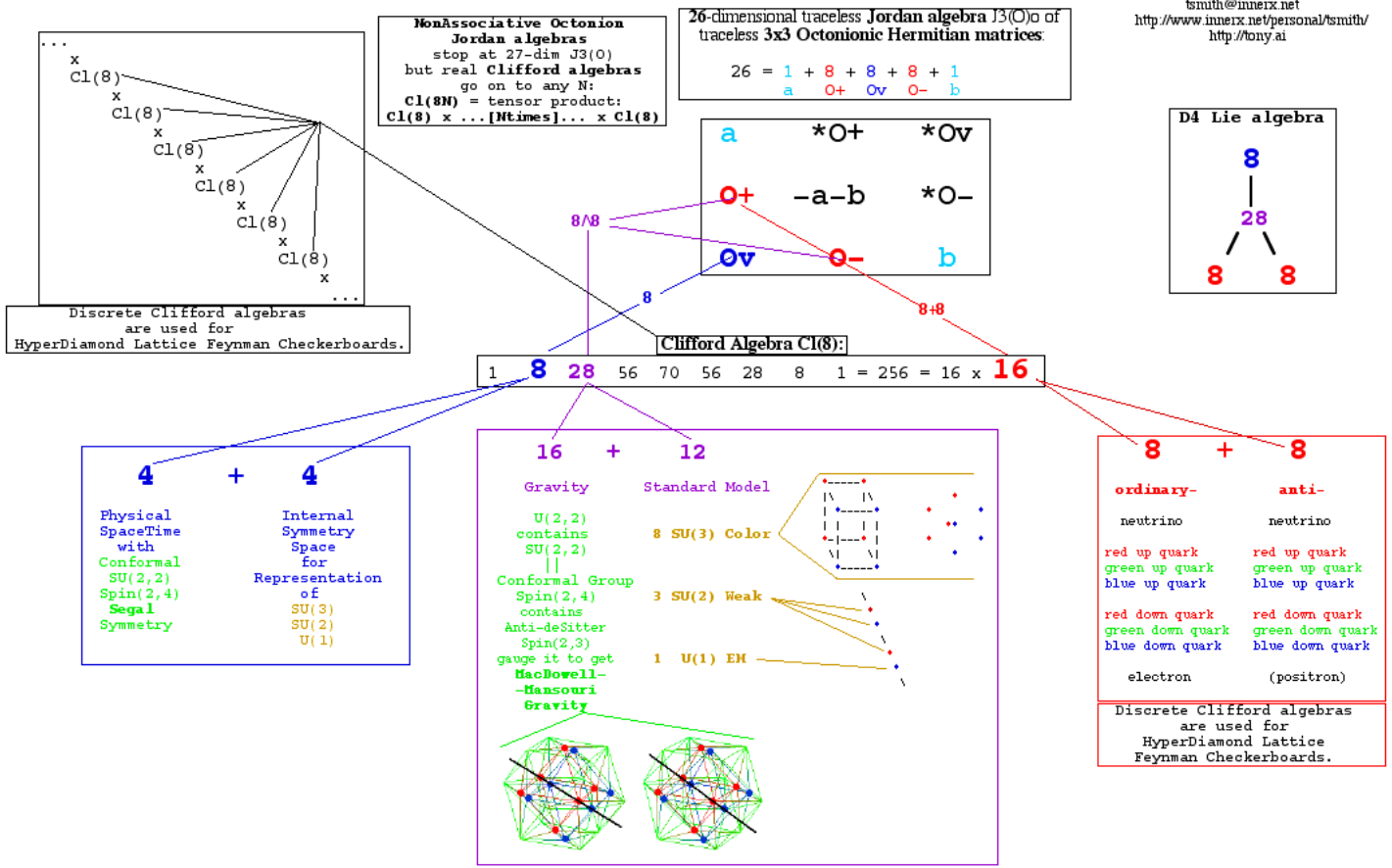

Note that:

- $27 = 12 + 15 = 6+6 + 6/\backslash 6$
- $27 = 8+8 + 8 + 3 = 6+6 + 2+2+8+3$

Details about some material mentioned on the above chart can beseen on these web pages:

- Clifford algebras - http://www.innerx.net/personal/tsmith/clfpq.html
    - Discrete - http://www.innerx.net/personal/tsmith/Sets2Quarks2.html#sub2
    - Real - http://www.innerx.net/personal/tsmith/clfpq.html#whatclifspin
- Octonions - http://www.innerx.net/personal/tsmith/3x3OctCnf.html
- Jordan algebras - http://www.innerx.net/personal/tsmith/Jordan.html

- Lie algebras - http://www.innerx.net/personal/tsmith/Lie.html
- Internal Symmetry Space - http://www.innerx.net/personal/tsmith/See.html
- Segal Conformal theory - http://www.innerx.net/personal/tsmith/SegalConf.html
- MacDowell-Mansouri gravity - http://www.innerx.net/personal/tsmith/cnfGrHg.html
- Standard Model Weylgroups - http://www.innerx.net/personal/tsmith/Sets2Quarks4a.html#WEYLdimredGB
- Fermions - http://www.innerx.net/personal/tsmith/Sets2Quarks9.html#sub13
- HyperDiamond lattices - http://www.innerx.net/personal/tsmith/HDFCmodel.html
- Generalized Feynman Checkerboards - http://www.innerx.net/personal/tsmith/Fynckb.html

---

The following sections are about:

- MacroSpace of Many-Worlds
- Unoriented Closed Bosonic Strings
- M-theory of the full 27-dimensional Jordan algebra J3(O)
  - F-theory of the 28-dimensional Jordan algebra J4(Q)
- Some descriptions of a few relevant terms
- 26-dimensional Bosonic Strings and the FakeMonster

To see some interesting connections among such things as the 24-dimensional Leech Lattice and the 256-dimensional Cl(8) Clifford Algebra, go to http://www.innerx.net/personal/tsmith/SegalConf2.html#MarkoRodin or click here.

---

The Lie algebra E6 of the D4-D5-E6-E7-E8 VoDou Physics model can be represented in terms of 3 copies of the 26-dimensional traceless subalgebra J3(O)o of the 27-dimensional Jordan algebra J3(O) by using the fibration E6 / F4 of 78-dimensional E6 over 52-dimensional F4 and the structure of F4 as doubled J3(O)o based on the 26-dimensional representation of F4. In this view, **the 26-dimensional traceless subalgebra J3(O)o is a representation of**

**the 26-dim Theory of Unoriented Closed Bosonic Strings produces a Bohm Quantum Theory with geometry of E6 / F4**.

In such an interpretation:

- the Real Shilov Boundary SpaceTime including Internal Symmetry Space and the Fermion Representation Space correspond to Pointlike States of a Point Particle Theory; and
- the Complex Dimensions of the Complex Domains of Complex Fermion Representation Space E6/ D5xU(1) and Complex SpaceTime related to D5 /D4xU(1) and D3 / D2xU(1) correspond to Stringlike States of a String Theory, the String being an extension of the Point Particle from the Real Shilov Boundary to its Complex Bounded Domain. Each Point of the Shilov Boundary extends to a Pencil of Parallel Lines in the Complex Bounded Domain. The String Theory of the Stringlike Parallel Lines produces the Bohm Quantum Theory. The set of all Pencils corresponds to the MacroSpace of Many-Worlds in a Many-Worlds Quantum Theory.

Click here to see how the Bosonic String Theory of 26-dim J3(O)o is related to an M-theory based on the full 27-dimensional J3(O).

Click here to see how the Bosonic String Theory of 26-dim J3(O)o is related to an F-theory based on the 28-dimensional J4(Q).

---

## Closed Unoriented Bosonic Strings:

Michio Kaku, in his books, Introduction to Superstrings and M-Theory (2nd ed) (Springer-Verlag 1999) and Strings, Conformal Fields, and M-Theory (2nd ed) (Springer-Verlag 2000) diagrams the Unoriented Closed Bosonic String spectrum:

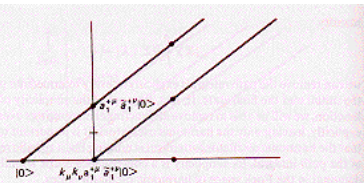

Figure 2.4. Regge trajectories for the closed string. The Fock space is built out of two commuting sets of harmonic oscillators. The massless spin-2 particle is the graviton, which corresponds to the product of both types of operators acting on the vacuum.

Joseph Polchinski, in his books String Theory vols. I and II (Cambridge 1998), says: "... [In] the simplest case of 26 flat dimensions ... the closed bosonic string ... theory has the maximal 26-dimensional Poincare invariance ... [and] ... is the unique theory with this symmetry ... It is possible to have a consistent theory with only closed strings ... **massless spectra**, with Guv representing the graviton [and] ... PHI the dilaton ... [and also] ... the tachyon ... [for the] **Closed unoriented bosonic string** [are]...":

- **massless spin-2 Gravitons** Guv, as to which Green, Schwartz, and Witten, in their book Superstring Theory, vol. 1, p.181 (Cambridge 1986) say "... the long-wavelength limit of the interactions of the massless modes of the bosonic closed string... [which] ... can be put in the form
  $$\text{INTEGRAL } d^{26} x \sqrt{g}\, R$$
  
  ...[of 26-dimensional general relativistic Einstein Gravitation]... by absorbing a suitable power of exp(-PHI) in the definition of the [26-dimensional MacroSpace] space-time metric g_uv ...";

- **scalar Dilatons** PHI, as to which Joseph Polchinski says "... The massless dilaton appears in the tree-level spectrum of every string theory, but not in nature: it would mediate a long-range scalar force of roughly gravitational strength. Measurements of the gravitational force at laboratory and greater scales restrict any force with a range greater than a few millimeters ( corresponding to a mass of order of 10^(-4) eV ) to be several orders of magnitude weaker than gravity, ruling out a massless dilaton. ...". In the D4-D5-E6-E7-E8 VoDou Physics model, Dilatons could **get an effectively real mass** through dimensional reduction of spacetime and through the X-scalar Higgs field of SU(5) GUT and the ElectroWeak SU(2)xU(1) Higgs scalar field and related conformal structures; and

- **Tachyons** with **imaginary mass**, as to which Joseph Polchinski says "... the negative mass-squared means that the no-string 'vacuum' is actually unstable ... whether the bosonic string has any stable vacuum ... the answer is not known. ...". In the interpretation of Closed Unoriented Bosonic String Theory as the MacroSpace of the Many-Worlds of World Strings, the instability of a no-string vacuum is natural, because:
    - if MacroSpace had no World Strings, or just one World String, the other possible World Strings would automatically be created, so that any MacroSpace would be "full" of "all" possible World Strings.

### What about the size/scale of each of the 26 dimensions of Closed Unoriented Bosonic String Theory?

Represent the size/scale of each dimension as a radius R, with R = infinity representing a flat large-scale dimension. Let $L_{pl}$ denote the Planck length, the size of the lattice spacing in the HyperDiamondLattice version of the D4-D5-E6-E7-E8 VoDou Physics model. Joseph Polchinski says "... as R -> infinity winding states become infinitely massive, while the compact momenta go over to a continuous spectrum. ... at the opposite limit R -> 0 ... the states with compact momentum become infinitely massive, but the spectrum of winding states ... approaches a continuum ... it does not cost much energy to wrap a string around a small circle. Thus as the radius goes to zero the spectrum again seems to approach that of a noncompact dimension. ... In fact, the R -> 0 and R -> infinity limits are physically identical. The spectrum is invariant under ...[

$$R \to R' = (L_{pl})^2 / R$$

]... This equivalence is known as T-duality. ... The space of inequivalent theories is the half-line [ $R \geq L_{pl}$ ]. We could take instead the range [ $0 \leq R \leq L_{pl}$ ] but it is more natural to think in terms of the larger of the two equivalent radii ... in particular questions of locality are clearer in the larger-R picture. Thus [from the larger-R point of view], there is no radius smaller than the self-dual radius [$R_{self-dual} = L_{pl}$ ]. ...". T-duality structures are is similar to Planck Pivot Vortex structures.

Consider a (purple) world-line String of one World of the MacroSpace of Many-Worlds and its interactions with another (gold) world-line World String, from the point of view of one point of the (purple) World String, seen so close-up that you don't see in the diagram that the (purple) and (gold) World Strings are both really closed strings when seen at very large scale:

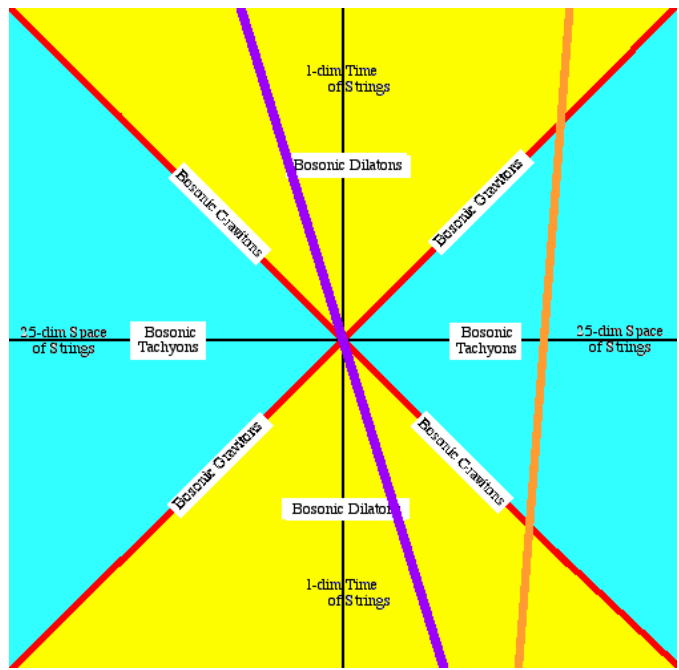

From the given point (diagram origin) of the (purple) World String:

- **massless spin-2 Gravitons** travel along the (red) MacroSpace light-cones to interact with the intersection points of those (red) light-cones with the (gold) World String;
- **scalar Dilatons, with effectively real mass,** travel within the (yellow) MacroSpace light-cone time-like interior to interact with the intersection region of the (yellow) light-cone time-like interior region with the (gold) World String; and
- **Tachyons, with imaginary mass,** travel within the (cyan) MacroSpace light-cone space-like exterior to interact with the intersection points of the (cyan) light-cone space-like exterior region with the (gold) WorldString.

In the D4-D5-E6-E7-E8 VoDou Physics model gravitation in the 26-dimensional Bosonic String Theory MacroSpace of the Many-Worlds justifies the Hameroff/Penrose idea:

**Superposition Separation is the separation/displacement of a mass separated from its superposed self. The picture is spacetime geometry separating from itself.**

[ Note that Gravity may not propagate in the 26 dimensions of the MacroSpace of the Many-Worlds in exactly the same way as it propagates in our 4-dimensional physical SpaceTime.]

**Gravitation from nearby World Strings might account for at least some Dark Matter that is indirectly observed in our World String**, an idea similar to one described (in the context of a superstring model that is in many ways very different from the D4-D5-E6-E7-E8 VoDou Physics model) by Nima Arkani-Hamed, Savas Dimopoulos, Gia Dvali, Nemanja Kaloper in their paper ManyfoldUniverse, hep-ph/9911386, and also in an article by the first three authors in the August 2000 issue of ScientificAmerican.

Bosonic Unoriented Closed String Theory describes the structure of Bohm's SuperImplicate Order MacroSpace and is related through (1+1) conformal structures to the LargeN limit of the AN Lie Algebras. For a nice introductory discussion of the mathematics of Bosonic Closed Strings, see Week 126 and Week 127 and other relevant works of John Baez.

**Branching among the Worlds of the Many-Worlds may be describable in terms of Singularities**, such as:

- simple singularities (classified precisely by the Coxeter groups Ak, Dk, E6, E7, E8);
- unimodal singularities (a single infinite three-suffix series and 14 "exceptional" one-parameter families ); and
- bimodal singularities ( 8 infinite series and 14 exceptional two-parameter families ).

---

### An M-theory of the full 27-dimensional Jordan algebra J3(O)

that could be S-dual to Bosonic String theory representing MacroSpace on 26-dim J3(O)o has been discussed in some recent (1997 and later) papers. In this model, **27-dimensional M-theory of bosonic strings has the geometry of E7 /E6xU(1).**

A physical interpretation of 27-dimensional J3(O) ( corresponding to J4(Q)o ) M-theory could be as a theory of Timelike Brane-Universes.

Timelike Brane-Universes might be considered as World-Lines (1-dimensional World-Lines with respect to the Shilov Boundary Pointlike States, but 1+1-dimensional Complex World-Lines (Complex Lines being like 2-dimensional Sheets or Membranes) with respect to the Complex Bounded Domain Stringlike States) in a Many-Worlds Quantum Theory.

In this view, each World of the MacroSpace of Many-Worlds can be seen as a 1-Timelike-dimensional String of Spacelike States, like a World Line or World String, and the **MacroSpace** of Many-Worlds can be represented geometrically by **E7/ E6xU(1) with 54 real dimensions and 27 complex dimensions** corresponding to the complexification of the 27-dimensional Jordan algebra J3(O) and algebraically by structure related to the same **27-dimensional Jordan algebra J3(O)**.

Discussing both open and closed bosonic strings, Soo-Jong Rey, in his paper hep-th/9704158, Heterotic M(atrix) Strings and Their Interactions, says:

"... We would like to conclude with a highly speculative remark on a possible **M(atrix) theory description of bosonic strings**. It is well-known that **bosonic Yang-Mills theory in twenty-six dimensions is rather special** ... The regularized one-loop effective action of d-dimensional Yang-Mills theory ...For d=26, the gauge kinetic term does not receive radiative correction at all ... We expect that this non-renormalization remains the same even after dimensional reductions. ... one may wonder if it is possible to construct M(atrix) string theory ... for bosonic string as well despite the absence of supersymmetry and BPS states.

The bosonic strings also have D-brane extended solitons ...whose tension scales as 1 / gB for weak string coupling gB<< 1. Given the observation that the leading order string effective action of **graviton, dilaton** and antisymmetric tensor field **may be derived from an Einstein gravity in d =27**, let us make an assumption that the 27-th `quantum'dimension decompactifies as the string coupling gB becomes large. For D0-brane, the dilaton exchange force may be interpreted as the 27-th diagonal component of d = 27 metric. Gravi-photon is suppressed by compactifying 27-th direction on an orbifold[ such as S1 / Z2 ] rather than on a circle. Likewise, its mass may be interpreted as 27-th Kaluza-Klein momentum of a massless excitation in d = 27. In the infinite boost limit, the light-front view of a bosonic string is that infinitely many D0-branes are threaded densely on the bosonic string. ...".

Gary T. Horowitz and Leonard Susskind, in their paper hep-th/0012037, Bosonic M Theory, say:

"... The possibility that **the bosonic string has a 27 dimensional origin** was ... discussed ...[ by Soo-Jong Rey in his paper hep-th/9704158 ]... in the context of a proposed matrix string formulation.... We conjecture that there exists a strong coupling limit of bosonic string theory which is related to the 26 dimensional theory in the same way that 11 dimensional M theory is related to superstring theory. More precisely, we believe that **bosonic string theory is the compactification on a line interval of a 27 dimensional theory whose low energy limit contains gravity and a three-form potential**. The line interval becomes infinite in the strong coupling limit, and this may provide a stable ground state of the theory. ...

we ... argue that the tachyon instability may be removed in this limit. ... The main clue motivating our guess comes from the existence of the dilaton and its connection to the coupling constant. ... Evidently, as in IIA string theory, the dilaton enters the action just as it would if it represented the compactification scale of a Kaluza Klein theory. We propose to take this seriously and try to interpret **bosonic string theory as a compactification of a 27 dimensional theory**. We will refer to this theory as **bosonic M theory** . ...

Closed bosonic string theory does not have a massless vector. This means it cannot be a compactification on an S1 . ...Accordingly, we propose that

**closed bosonic string theory is a compactification of 27 dimensional bosonic M theory on [an orbifold ] S1 / Z2**. ...

In the bosonic case, since there are no fermions or chiral bosons, there are no anomalies to cancel. So there are no extra degrees of freedom living at the fixed points. ... the weakly coupled string theory is the limit in which the compactification length scale becomes much smaller than the 27 dimensional Planck length and the strong coupling limit is the decompactification limit. The 27 dimensional theory should contain membranes but no strings, and would not have a dilaton or variable coupling strength. The usual bosonic string corresponds to a membrane stretched across the compactification interval. ... **the low energy limit of bosonic M theory ... is a gravity theory in 27 dimensions** ... In order to reproduce the known spectrum of weakly coupled bosonic string theory, bosonic M theory will have to contain an additional field besides the 27 dimensional gravitational field, namely a three-form potential CFT. Let us consider **the various massless fields that would survive in the weak coupling limit**.

- First of all, there would be the **26 dimensional graviton**. As usual, general covariance in 26 dimensions would insure that it remains massless.
- The component of the 27 dimensional gravitational field $g_{27,27}$ is a **scalar in the 26 dimensional theory**. It is of course the **dilaton**. No symmetry protects the mass of the dilaton. In fact we know that at the one loop level a dilaton potential is generated that lifts the dilatonic at direction. Why the mass vanishes in the weak coupling limit is not clear.
- Massless vectors have no reason to exist since there is no translation symmetry of the compactification space. This is obvious if we think of this space [ the

- orbifold S1 / Z2 ] as a line interval.
- ...[ with respect to **tachyons** ]... Even if 27 dimensional flat space, M27, is a stable vacuum, one might ask what is the "ground state" of the theory at finite string coupling, or finite compactification size? Tachyon condensation is not likely to lead back to M27, and there is probably no stable minimum of the tachyon potential in 26 dimensions ...Instead, we believe **tachyon condensation may lead to an exotic state with zero metric guv = 0**. It is an old idea that quantum gravity may have an essentially topological phase with no metric. We have argued that the tachyon instability is related to nucleation of "bubbles of nothing" which is certainly reminiscent of zero metric.

... As an aside, we note that there is also a brane solution of 26 dimensional bosonic string theory which has both electric and magnetic charge associated with the three-form H. It is a 21-brane with fundamental strings lying in it and smeared over the remaining 20 directions. Dimensionally reducing to six dimensions by compactifying on a small T 20 , one recovers the usual selfdual black string in six dimensions. ...

... We have proposed that a **bosonic version of M theory exists, which is a 27 dimensional theory with 2-branes and 21-branes**. One recovers the usual bosonic string by compactifying on S1 / Z2 and shrinking its size to zero. In particular, a Planck tension 2-brane stretched along the compact direction has the right tension to be a fundamental string. This picture offers a plausible explanation of the tachyon instability and suggests that uncompactified 27 dimensional flat space may be stable. A definite prediction of this theory is **the existence of a 2+1 CFT with SO(24) global symmetry**, which should be its holographic dual for AdS4 x S23 boundary conditions. ... if there does not exist a 2+1 CFT with SO(24) global symmetry, bosonic M theory would be disproven.

... **What kind of theory do we get if we compactify bosonic M theory on a circle instead of [ theorbifold S1 / Z2 ] a line interval**? ... we believe the limit of bosonic M theory compactified on a circle as the radius R --> 0 is the same as the limit R --> infinity, i.e., the uncompactified 27 dimensional theory. If we compactify bosonic M theory on S1 x ( S1 / Z2 ), and take the second factor very small, this is a consequence of the usual T-duality of the bosonic string. More generally, it appears to be the only possibility with the right massless spectrum. ...".

Lee Smolin, in hispaper The exceptional Jordan algebra and the matrix string, hep-th/0104050, says: "A new matrix model is described, based on the exceptional Jordan algebra, J3(O). The action is cubic, as in matrix Chern-Simons theory. We describe a compactification that, we argue, reproduces, at the one loop level, an octonionic compactification of the matrix string theory in which SO(8) is broken to G2. There are 27 matrix degrees of freedom, which under Spin(8) transform as the vector, spinor and conjugate spinor, plus three singlets, which represent the two longitudinal coordinates plus an eleventh coordinate. Supersymmetry appears to be related to triality of the representations of Spin(8).".

 YuhiOhwashi, in his paper E6 Matrix Model, hep-th/0110106, says: "... Lee Smolin's talk presented at The 10th Tohwa University International Symposium (July 3-7, 2001, Fukuoka, Japan) was my motive for starting this work. ...

... Smolin's matrix model [is] based on the groups of type F4. ... The action of Smolin's model is given ...[in terms of]... elements of exceptional Jordan algebra J..... The exceptional Jordan algebra J is a 27-dimensional R-vector space. This space can be classified into three main parts.

- One is the Jordan algebra j which is a 10-dimensional R-vector space. [ These 10 dimensions correspond in my D4-D5-E6-E7-E8 VoDou Physics model to 4-dimensional Physical Spacetime and 4-dimensional Internal Symmetry Space plus 2 extra dimensions. ]
- The others are the part of 16 dimensions which is related to the spinors [ These 10 dimensions correspond in my D4-D5-E6-E7-E8 VoDou Physics model to representation space for 8 first-generation fermion particles plus 8 first-generation fermion anti-particles. ]
- and the extra 1 dimension. ...

 ... If the standard model were described by using Majorana spinors only, F4 might be the underlying symmetry of the universe.... However, the actual world requires complex fermions without doubt. This is the reason why we have to abandon the simply connected compact exceptional Lie group F4. ... In accordance with this complexification, the groups of type F4 are upgraded to the groups of type E6. ...

... we consider a new matrix model based on the simply connected compact exceptional Lie group E6 ... The action of the model is constructed from cubic form which is the invariant on E6 mapping.

This action is an essentially complex action. Of course if one wants, one may take up only real part of that ...

Our model has twice as many degrees of freedom as Smolin's model has because we are considering E6 instead of F4. ... This is a future problem which needs to be asked. ...".

The ideas of Smolin and Ohwashi are related to my D4-D5-E6-E7-E8 VoDou Physics model in interesting ways:

- both Smolin and I came to 27-dimensional M-theory through the paper hep-th/0012037 by Horowitz and Susskind;
- both Smolin and the D4-D5-E6-E7-E8 VoDou Physics model use the Jordan algebra structure of J3(O) and its relationship to the three triality representations of Spin(8);
- with respect to supersymmetry (although details may differ somewhat) our approaches to supersymmetry are similarly motivated:
  - Smolin approaches supersymmetry "... as related to that part of the F4 algebra that is generated by Spin(8) spinorial variables ..."; and
  - the approach of the D4-D5-E6-E7-E8 VoDou Physics model is
    - to identify the two 8-dimensional half-spinor representations of Spin(8) with the 8 first generation particles and the 8 corresponding antiparticles,

      each of which are then identified by Spin(8) triality with

    - the 8-dimensional vector representation of Spin(8), which then produces by the wedge product the 8 /\ 8 = 28 gauge bosons of the bivector adjoint representation of Spin(8), and then reduces by dimensional reduction of 8-dimensional spacetime to 4-dimensional phyiscal spacetime plus 4-dimensional internal symmtery space to
      - 16 U(2,2) bosons of MacDowell-Mansouri gravity plus the Higgs mechanism plus a propagator phase
      - plus the 12 bosons of the SU(3)xSU(2)xU(1) Standard Model,

    leading to a subtletriality supersymmetry;
- both Ohwashi and I decided that an E6 model would be better than an F4 model. Here is some history of my progression from F4 to E6.

However, there are differences between the approaches of Smolin and Ohwashi, and the approach of the D4-D5-E6-E7-E8 VoDou Physics model. For example:

- They use a cubic action related to Chern-Simons theory, while the action of the D4-D5-E6-E7-E8 VoDou Physics model is fundamentally an 8-dimensional Lagrangian that produces MacDowell-Mansouri Gravity, the Higgs Mechanism, and the Standard Model upon dimensional reduction of SpaceTime; and
- I solve the "twice as many degrees of freedom" problem of complex structures associated with E6 by using Shilov boundaries to represent the physically effective parts of complex structures. Since E6/ D5xU(1) is not a tube-type domain, its Shilov boundary is not totally real, so I regard its real part as representing the 8 first-generation fermion particles, with the anti-particles being an imaginary part.

Metod Saniga, in physics/0012033, discusses in the context of string theory (although in a different context ( heterotic superstrings ) from that of the D4-D5-E6-E7-E8 VoDou Physics model ) another 27-dimensional structure, saying:

> "... It is a well-known fact that on a generic cubic surface, K3, there is a configuration of twenty-seven lines ... the lines are seen to form three separate groups. The first two groups, each comprising six lines, are known as Schlafli's double-six. The third group consists of fifteen lines ... The basics of the algebra can simply be expressed as 27 = 12 + 15 ...".

It is interesting to contemplate the relationship between the 3x3 matrix structure

```
         1     8     8
         -     1     8
         -     -     1
```

of the 27-dimensional Jordan algebra J3(O) and the 27-line geometry structure

```
         6+6  +  6/\6  =  6+6  +  15  =  27
```

Let the 8 be represented by 8-dimensional octonions, with basis {1,i,j,k,E,I,J,K}, and let the 6 be represented by a 6-dimensional subspace, with basis {i,j,k,I,J,K}. Let the two 6s of 6+6 be represented as subspaces of the two next-to-diagonal 8s of the J3(O) matrix:

```
         -     6     -
         -     -     6
         -     -     -
```

then the 6/\6 = 15 lines of the 27-line might correspond to the

```
         1     2     8
         -     1     2
         -     -     1
```

in terms of the J3(O) matrix. Here are some more relevant relationships:

- the Symmetry Group of the 27-line is the Weyl Group of the Lie Algebra E6;
- the Lie Group E6 is the Automorphism Group of the 56-dimensional Freudenthal Algebra Fr3(O), which can be visualized as a complexification of the 27-dimensional Jordan Algebra J3(O);
- the Symmetric Space E7 /E6xU(1) can be visualized as a complexification of the 27-dimensional Jordan Algebra J3(O), and its Shilov Boundary

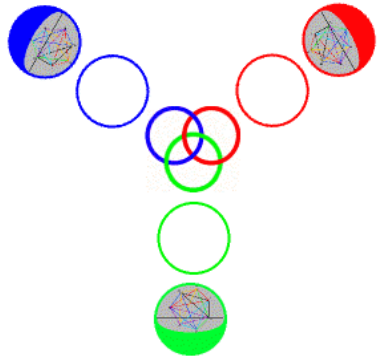

  as (1+26)-dimensional S1x J3(O)o, which is very similar to 27-dimensional J3(O) itself;

- the Automorphism Group of the 27-dimensional Jordan Algebra J3(O) ( which is sometimes also denoted H3(O) ) is the Lie Group F4;
- the Lie Algebra F4 has D4-B4-F4 structure that can be visualized as a real version of the D4-D5-E6 structure of the D4-D5-E6-E7-E8 VoDou Physics model.

Metod Saniga's ideas have been referenced by Carlos Castro in physics/0104016, in which Carlos Castro says:

> "... Motivated by the fact that the bosonic membrane is devoid of anomalies in d = 27, and the supermembrane is anomaly free in d = 11, and that the anomaly free ( super) string actions ( d = 26, 10 ) are directly obtained by a double-dimensional reduction process of both the world-volume of the ( super)membrane and the target spacetime dimension, where the (super)membrane is embedded, we shall derive rigorously the transfinite M theoretical corrections ... to El Naschie's inverse fine structure constant ... which were based on a transfinite perturbative Heterotic string theory formalism ...".

Although Carlos Castro uses some similar mathematical structures, such as Clifford algebra, his physics model is different from the D4-D5-E6-E7-E8 VoDou Physics model in a number of respects, such as, particularly, his use of conventional superstring theory instead of using the D4-D5-E6-E7-E8 VoDou Physics model viewpoint of seeing bosonic strings as WorldLine Worlds in the MacroSpace of the Worlds of the Many-Worlds.

However, some of the interesting similarities that I perceive include:

- some of Carlos Castro's ideas about Prime Numbers are related to the role of Prime Numbers in the MacroSpace of the Many-Worlds, and that
- some of his ideas about calculation of such things of the ElectroMagnetic Fine Structure Constant are related to the fact that the geometry of the MacroSpace of the Many-Worlds is closely related to the geometry of the Forces and Particles of the D4-D5-E6-E7-E8 VoDou Physics model, so that there are similarities among all the levels of structures of the D4-D5-E6-E7-E8 VoDou Physics model,

**Branching among the Worlds of 27-dim M-Theory may be describable in terms of Singularities**, such as:

- simple singularities (classified precisely by the Coxeter groups Ak, Dk, E6, E7, E8);
- unimodal singularities ( a single infinite three-suffix series and 14 "exceptional" one-parameter families ); and
- bimodal singularities ( 8 infinite series and 14 exceptional two-parameter families ).

In bosonic string theory, 27-dimensional M-theory is a subspace of

## 28-dimensional F-theory with Jordan algebra J4(Q).

A physical interpretation of 28-dimensional J4(Q) F-theory could be as a theory of Spacelike Brane-Universes.

Spacelike Brane-Universes might be considered as Spatial Worlds (3-dimensional Spatial Worlds with respect to the Shilov Boundary Pointlike States, but 3+3-dimensional Complex Spatial Worlds with respect to the Complex Bounded Domain Stringlike States) in a Many-Worlds Quantum Theory.

**28-dimensional F-theory of bosonic strings has the geometry of E8 /E7xSU(2).**

Bosonic string F-theory is described by Jose M Figueroa-O'Farrill, in his paper F-theory and the universal string theory, hep-th/9704009:

> "... Let us first consider a bosonic string background.... The graviton couples to the energy-momentum tensor T. If we now add a U(1) gauge field, it will couple to a vector J. We therefore would like to investigate under which conditions the algebra generated by T and J can be used consistently to define a (generalised) string theory. ...
>
> ... A particular realisation of this algebra is provided by a **bosonic string propagating in a 28-dimensional pseudo-euclidean space with signature ( 26, 2 )**. The signature can be understood from unitarity ...
>
> ... the BRST cohomology of this system agrees with that of **an underlying bosonic string propagating in a 26-dimensional Minkowski subspace** perpendicular to v and not containing v, provided that we identify states whose momenta differ by a multiple of v ...
>
> ... Suppose that T is the energy-momentum tensor of a critical **bosonic string propagating in 26-dimensional Minkowski spacetime**.
>
> Then T corresponds to **the string propagating on a ( 26 +2)-dimensional pseudo-euclidean space**.
>
> The BRST operator is invariant under the subgroup of the ( 26 +2 ) pseudo-euclidean group of motions which preserves the null vector v. This is nothing but the ( 25 + 1 ) conformal group, which does not act linearly in Minkowski spacetime but does on the larger space.
>
> Symmetries of the BRST operator induce symmetries in the cohomology, hence we would expect that the spectrum should assemble itself into representations on the conformal group. We know that the physical spectrum of the bosonic string only possesses ( 25 + 1 ) Poincare covariance, so what happens to the special conformal transformations?
>
> ... bosonic ghosts ... have a (countably) infinite number of inequivalent vacua which can be understood as the momenta in one of two auxiliary compactified dimensions introduced by the bosonisation procedure. The picture changing operator interpolates between these different vacua, commuting with the BRST operator and thus introducing an infinite degeneracy in the cohomology....
>
> .. the special conformal transformations ... change the picture. By definition a picture-changing operator is a BRST invariant operator which changes the picture, whence the special conformal transformations are picture-changing operators.
>
> A remarkable fact of this treatment is that the appearance of the lorentzian torus is very natural. In other words, by enhancing the gauge principle on the worldsheet to incorporate the extra U(1) gauge invariance we are forced to reinterpret bosonic string vacua corresponding to propagation on a given manifold M, as propagation in a manifold which at least locally is of the form Mx T 2 where T 2 is the lorentzian torus corresponding to the bosons ... **This theory is precisely the F-theory** introduced...[ by Vafa in hep-th/9602022 ]... except that there the compactness of the extra two coordinates was an ad hoc assumption. ...".

**Branching among the Worlds of the Spacelike Branes of 28-dim F-Theory may be describable in terms of Singularities**, such as:

- simple singularities (classified precisely by the Coxeter groups Ak, Dk, E6, E7, E8);
- unimodal singularities ( a single infinite three-suffix series and 14 "exceptional" one-parameter families ); and
- bimodal singularities ( 8 infinite series and 14 exceptional two-parameter families ).

---

**Here are some descriptions of a few relevant terms:**

Michio Kaku, in his book Introduction to Strings and M-Theory (second edition, Springer 1999), says:

> "... the **closed [super] string ( Type II )** ... the fields can either be chiral or not. Closed strings are, by definition, periodic in sigma, which yields the following normal mode expansion:
>
> - S1a(s,t) = Sum( n = -infinity; n = + infinity ) San exp( -2 i n( t - s ) ) ,
> - S2a(s,t) = Sum( n = -infinity; n = + infinity ) S'an exp( -2 i n( t + s ) ) .
>
> If these two fields have different chiralities, then they are called **Type IIA**. ... this **represents the N = 2, D =10-dimensional reduction of ordinary N = 1, D = 11 supergravity**. ...
>
> ... there exists a new 11-dimensional theory, called **M-theory**, containing 11-dimensional supergravity as its low-energy limit, which reduces to Type IIA [super] string theory (with Kaluza-Klein modes) when compactified on a circle. ... the strong coupling limit of 10-dimensional Type IIA superstring theory is equivalent to the weak coupling limit of a new 11-dimensional theory [ M-theory ], whose low-energy limit is given by 11-dimensional supergravity. ... Using perturbation theory around weak coupling in 10-dimensional Type IIA superstring theory, we would never see 11-dimensional physics, which belongs to the strong coupling region of the theory. ... M-theory is much richer in its structure than string theory. In M-theory, there is a three-form field Amnp, which can couple to an extended object. We recall that in electrodynamics, a point particle acts as the source of a vector field Au. In [open] string theory, the [open] string acts as the source for a tensor field Buv. Likewise, in M-theory, a membrane is the source for Amnp. ...
>
> ... Ironically, 11-dimensional supergravity was previously rejected as a physical theory because:
>
> - (a) it was probably nonrenormalizable (i.e., there exists a counterterm at the seventh loop level);
> - (b) it does not possess chiral fields when compactified on manifolds; and
> - (c) it could not reproduce the Standard Model, because it could only yield SO(8) when compactified down to four dimensions.
>
> Now we can veiw 11-dimensional supergravity in an entirely new light, as the low-energy sector of a new 11-dimensional theory, called M-theory, which suffers from none of these problems. The question of renormalizability is answered because the full M-theory apparently has higher terms in the curvature tensor which render the theory finite. The question of chirality is solved because ... M-theory gives us chirality when we compactify on a space which is not a manifold (such as [ orbifolds such as S1 / Z2 ] line segments). And the problem that SO(8) is too small to accommodate the Standard Model is solved when we analyze the theory nonperturbatively, where we find E8 xE8 symmetry emerging when we compactify on [ orbifolds such as S1 / Z2 ] line segments. ...".
>
>> Note that the D4-D5-E6-E7-E8 VoDou Physics Model solves the problems of 11-dimensional supergravity in different ways, but uses many similar mathematical structures and techniques.

Michio Kaku, in his book Strings, Conformal Fields and M-Theory (second edition, Springer 2000), says:

> "... **S: M-theory on S1 <---> IIA** ... Type IIA [super] string theory is S dual to a new, D = 11 theory called M-theory, whose lowest-order term is given by D = 11 supergravity. ...
>
> ... **S: M-theory on S1 / Z2 <---> E8 x E8** ...[11-dimensional ]... M-theory, when compactified on a line segment [S1 / Z2 ], is dual to the ... [ E8 x E8 heterotic ]... string ...".

Lisa Randall and Raman Sundrum, in their paper hep-ph/9905221, say:

> "... we work on the space **S1 / Z2**. We take the range of PHI to be from -pi to pi; however the metic is completely specified by the values in the range 0 $\leq$ PHI $\leq$ pi. The **orbifold** fixed points at PHI = 0, pi...[may]... be taken as the locations of ... branes ...".
>
> ```
>        Note that S1 / Z2 can have two different interpretations.
>              John Baez says:
>        "... Z_2 acts in various ways on the circle.
>        Let's think of the circle as the subset
>        {(x,y): x^2 + y^2 = 1}  of R^2.
>        Z_2 can act on it like this:
>        (x,y) |-> (-x,-y)
>        and then S^1/Z_2 = RP1 [Real Projective 1-space]
>        which is a manifold, in fact a circle.
>        ...
>        Z_2 also can act on the circle like this:
>        (x,y) |-> (-x,y)
>        and then S^1/Z_2 is an orbifold,
>        in fact a closed interval. ...".
>
>        The physical interpretations of RP1 in
>        the D4-D5-E6-E7-E8 VoDou Physics model
>        as Time of SpaceTime and
>        as representation space for Neutrino-type
>        (only one helicity state) Fermions
>        might be viewed as having some
>        of the characteristics of a orbifold line interval.
> ```

Joseph Polchinski, in his book String Theory (volume 1, Cambridge 1998), says:

> "... **orbifold**
>
> - 1. ... **a coset space M / H** , where H is a group of discrete symmetries of a manifold M. **The coset is singular at the fixed points of H** ;
> - 2. ... the CFT or string theory produced by the gauging of a discrete world-sheet symmetry goup H. If the elements of H are spacetime symmetries, the result is a theory of strings propagating on the coset space M / H . A non-Abelian orbifold is one whose point group is non-Abelian. An asymmetric orbifold is one where H does not have a spacetime interpretation and which in general acts differently on the right-movers and left-movers of the string;
> - 3. ... to produce such a CFT or string theory by gauging H; this is synonymous with the second definitioin of twist.

> ... **S-duality** ... a duality under which the coupling constant of a quantum theory changes nontrivially, including the case of weak-strong duality. ... In compactified theories, the term S-duality is limited to those dualities that leave the radii invariant, up to an overall coupling-dependent rescaling ...
>
> ... **T-duality** ... a duality in string theory, usually in a toroidally compactified theory, that leaves the coupling constant invariant up to a radius-dependent rescaling and therefore holds at each order of string perturbation theory. Most notable is R --> a' / R duality, which relates string theories compactified on large and small tori by interchanging winding and Kaluza-Klein states. ...
>
> ... **U-duality** ... any of the dualities of a string theory ... This includes the S-dualities and T-dualities, but in contrast to these includes also transformations that mix the radii and couplings. ...".

---

### 26-dimensional Bosonic Strings and the Fake Monster

R. E. Borcherds, in his paper Problems in Moonshine, says: "... The **classification of finite simple groups** shows that every finite simple group either fits into one of about 20 infinite families, or is one of **26 exceptions, called sporadic simple groups**. The **monster simple group is the largest of the sporadic finite simple groups**, and was discovered by Fischer and Griess ... Its order is

8080, 17424, 79451, 28758, 86459, 90496, 17107, 57005, 75436, 80000, 00000

=

$2^{46} . 3^{20} . 5^9 . 7^6 . 11^2 . 13^3 . 17 . 19 . 23 . 29 . 31 . 41 . 47 . 59 . 71$

(which is roughly [ $8 \times 10^{53}$ ] the number of elementary particles in the earth [ actually, the earth's mass is about $6 \times 10^{51}$ GeV, and it is Saturn that has mass about $6 \times 10^{53}$ GeV, or about $6 \times 10^{53}$ hydrogen masses ]). The smallest irreducible representations have dimensions 1, 196883, 21296876, . . ..

On the other hand the elliptic modular function j(t) ... has the power series expansion

$j(t) = q^{(-1)} + 744 + 196,884\ q + + 21,493,760 q\ 2 + ...$

where q = exp( 2 pi i t ). John McKay noticed some rather weird relations between coefficients of the elliptic modular function and the representations of the monster as follows:

- 1 = 1
- 196884 = 196883 + 1
- 21493760 = 21296876 + 196883 + 1

where the numbers on the left are coefficients of j(t) and the numbers on the right are dimensions of irreducible representations of the monster. The term "monstrous moonshine" (coined by Conway) refers to various extensions of McKay's observation, and in particular to relations between sporadic simple groups and modular functions....

Allcock ... recently constructed some striking examples of complex hyperbolic reflection groups from the Leech lattice, or more precisely from the complex Leech lattice, a 12 dimensional lattice over the Eisenstein integers. This complex reflection group looks similar in several ways to Conway's real hyperbolic reflection group of the lattice /\25,1 ...".

Gregory Moore, in his paper Finite in All Directions, hep-th/9305139, says:

> "... At a generic point [ g of a string theory toroidal compactification lattice, the Lie algebra ] Lg = $IR^{26} + IR^{26}$ ...
>
> ... The distinguished point g* may be regarded as a point of maximal symmetry in the moduli space of toroidal compactifications...
>
> ... Given a point of maximal symmetry it is natural to ask if L* = Lg* is a universal symmetry of string theory in the sense that all other unbroken symmetry algebras which arise in toroidal compactification are subalgebras of L*. Unfortunately, maximal symmetry does not imply that L* is universal. ...
>
> ... For the bosonic string the Lie algebra L* is related to the Monster group. L* = A x A where A is the "Fake Monster Lie algebra" studied by Borcherds ...".

R. E. Borcherds, J. H. Conway, L. Queen and N. J. A. Sloane, in their paper A Monster Lie Algebra?, say: "... [A version of this paper was originally published in Advances in Mathematics, vol.53(1984), no. 1, pp. 75{79. A revised version appeared as Chapter 30 of "Sphere packing, lattices and groups" by J. H. Conway and N. J. A. Sloane, Springer-Verlag, 1988.] ... Remark added 1998:

> The Lie algebra of this paper is indeed closely related to the monster simple group. In order to get a well behaved Lie algebra it turns out to be necessary to add some imaginary simple roots to the "Leech roots". This gives the fake monster Lie algebra, which contains the Lie algebra of this paper as a large subalgebra.
>
> See R. E. Borcherds, "The monster Lie algebra", Adv. Math. Vol. 83, No.1, Sept. 1990, for details (but note that the fake monster Lie algebra is called the monster Lie algebra in this paper). The term "monster Lie algebra" is now used to refer to a certain \"=2Z-twisted" version of the fake monster Lie algebra. The monster Lie algebra is acted on by the monster simple group, and can be used to show that the monster module constructed by Frenkel, Lepowsky, and Meurman satisfies the moonshine conjectures; see R.E. Borcherds, "Monstrous moonshine and monstrous Lie superalgebras", Invent. Math. 109, 405-444 (1992). ...". Also, see the web seminar What is Moonshine?, Richard Borcherds, 25 November 1998. and the paper What is Moonshine?, math.QA/9809110.

Reinhold W. Gebert, in his paper Introduction to Vertex Algebras, Borcherds Algebras, and the Monster Lie Algebra, hep-th/9308151, says:

"... Borcherds algebras arise as certain "physical" subspaces of vertex algebras ... As a class of concrete examples the vertex algebras associated with even lattices are constructed and it is shown in detail how affine Lie algebras and the fake Monster Lie algebra naturally appear. This leads us to the abstract definition of Borcherds algebras as generalized Kac-Moody algebras and their basic properties. Finally, the results about the simplest generic Borcherds algebras are analysed from the point of view of symmetry in quantum theory and the construction of the Monster Lie algebra is sketched. ...

the fake Monster Lie algebra seemingly plays an important role in bosonic string theory ... Vertex algebras associated with even lattices have their origin in toroidal compactifications of bosonic strings. ... As an easy application we demonstrate how affine Lie algebras arise in this context. Furthermore, the fake Monster Lie algebra which is the first generic example of a Borcherds algebra, is worked out in detail. ...

the so called Moonshine Module ... constructed by Frenkel, Lepowsky, and Meurman is a vertex operator algebra ... and it turns out that ...[its]... weight two piece ... is a non-associative algebra with symmetric product ... and associative bilinear form ...[that]... is precisely the 196,884-dimensional Griess algebra which possesses the Monster group ... as its full automorphism group...

Things become more complicated when we move away from the lattice /\ being Euclidian. Let us consider the unique 26-dimensional even unimodular Lorentzian lattice /\25,1 ... In physics this corresponds to an open bosonic string moving in 26-dimensional spacetime compactified on a torus so that the momenta lie on a lattice. Calculations in connection with the automorphism group of /\25,1 show that ... [the] simple roots generate the reflection group of /\25,1 ... We shall also call the positive norm simple roots of /\25,1 Leech roots since Conway has shown that this subset is indeed isometric to the Leech lattice, the unique 24-dimensional even unimodular Euclidian lattice with no vectors of square length two. ... We now define a Kac-Moody algebra Linfinity, of infinite dimension and rank ...Linfinity has three generators ... for each Leech root ...

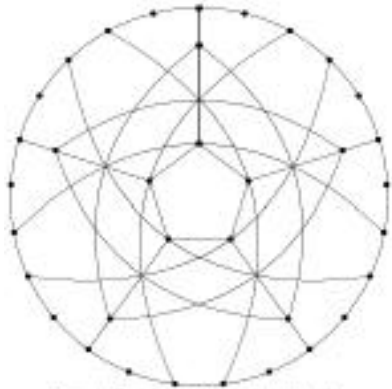

Figure 1: Part of the Dynkin diagram of L...

... Let us summarize: We define the fake Monster Lie algebra g/\25,1 to be the Lie algebra with root lattice /\25,1, whose simple roots are the simple roots of the Kac-Moody algebra Linfinity, together with the positive integer multiples of the Weyl vector ... each with multiplicity 24. ... the fake Monster Lie algebra is not a Kac-Moody algebra due to the presence of the lightlike simple Weyl roots which violate an axiom for these algebras ... Nevertheless, the structure of g/\25,1 resembles a Kac-Moody algebra very well. ...

Frenkel, Lepowsky and Meurman constructed the Monster vertex algebra which is acted on by the Monster simple sporadic group. The underlying vector space which is called Moonshine Module...[It]... provides a natural infinite-dimensional representation of the Monster [and it] is characterized by the following properties:

- (i) ...[It]... is a vertex operator algebra with a conformal vector ... of dimension 24 and a positive definite bilinear form
- (ii) ...[It is the sum of eigenspaces of L(0) with eigenvalues n+1 and with dimension ]... given via the generating function ...

$$[ \text{SUM}(n \geq -1) \dim(F_n) q^n = J(q) = j(q) - 744 = q^{-1} + 196{,}884\, q + ... ]...$$

- (iii) The Monster group acts on ...[it]...preserving the vertex operator algebra structure, the conformal vector ... and the bilinear form. ...

The Monster vertex algebra is realized explicitly as

$$F = F^+_{/\Lambda\text{Leech}+} + F^+_{/\Lambda\text{Leech}}$$

where F/\Leech denotes the vertex operator algebra associated with the Leech lattice, the unique 24-dimensional even unimodular Euclidian lattice with no elements of square length two. ... the Monster module [can be seen] as Z2-orbifold of a bosonic string theory compactified to the Leech lattice ...[and as]... a Zp-orbifold ... It is interesting that there is also an approach to the Monster module based on twisting the heterotic string. ...

The starting point for the definition of a Monster Lie algebra should be the fake Monster Lie algebra. We use the fact that the Lorentzian lattice /\25.1 can be written as the direct sum of the Leech lattice and the unique two-dimensional even unimodular Lorentzian lattice /\1,1. ... the vertex algebra associated with the Lorentzian lattice /\25,1 is the tensor product of the vertex algebras corresponding to F/\Leech and F/\1,1 . One finds that the Leech lattice gives rise to a vertex operator algebra with conformal vector of dimension 24 and a positive definite bilinear form. Furthermore, F/\Leech = SUM(n$\geq$1) Fn/\Leech where Fn/\Leech ... is the eigenspace of L(0) with eigenvalue n+1 and the dimension of Fn/\Leech is given via the generating function...

$$\text{SUM}(n \geq 1) \dim(F_n/\Lambda\text{Leech}) q^n = J(q) = j(q) - 720 = q^{-1} + 24 + 196{,}884\, q + ...$$

... the Monster Lie algebra is seen to be a generalized Kac-Moody algebra. ...".